\pdfoutput=1
\documentclass[journal=mamobx,manuscript=article]{achemso}

\usepackage[version=3]{mhchem} 
\usepackage{amssymb}
\usepackage[usenames,dvipsnames]{xcolor}

\author{Rodrigo Rivas-Barbosa}
\affiliation{Department of Physics, Sapienza University of Rome, Piazzale Aldo Moro 2, 00185 Roma, Italy}
\alsoaffiliation{Divisi{\'o}n de Ciencias e Ingenier{\'i}as, Universidad de Guanajuato, Lomas del Bosque 103, 37150 Le{\'o}n, Mexico}
\altaffiliation{These authors contributed equally to this work}
\author{Jos{\'e} Ruiz-Franco}
\affiliation{Department of Physics, Sapienza University of Rome, Piazzale Aldo Moro 2, 00185 Roma, Italy}
\alsoaffiliation{CNR Institute of Complex Systems, Uos Sapienza, Piazzale Aldo Moro 2, 00185, Roma, Italy}
\alsoaffiliation{Physical Chemistry and Soft Matter, Wageningen University \& Research,\\ Stippeneng 4, 6708WE Wageningen, The Netherlands}
\altaffiliation{These authors contributed equally to this work}
\author{Mayra A. Lara-Pe{\~n}a}
\affiliation{Divisi{\'o}n de Ciencias e Ingenier{\'i}as, Universidad de Guanajuato, Lomas del Bosque 103, 37150 Le{\'o}n, Mexico}
\author{Jacopo Cardellini}
\affiliation{Dipartimento di Chimica and CSGI, Università di Firenze, 50019 Sesto Fiorentino, Italy}
\author{Angel Licea-Claverie}
\affiliation{Centro de Graduados e Investigaci{\'o}n en Química del Tecnol{\'o}gico Nacional de M{\'e}xico/Instituto Tecnol{\'o}gico de Tijuana, 22500 Tijuana, Mexico}
\author{Fabrizio Camerin}
\affiliation{CNR Institute of Complex Systems, Uos Sapienza, Piazzale Aldo Moro 2, 00185, Roma, Italy}
\alsoaffiliation{Department of Physics, Sapienza University of Rome, Piazzale Aldo Moro 2, 00185 Roma, Italy}
\author{Emanuela Zaccarelli}
\email{emanuela.zaccarelli@cnr.it}
\affiliation{CNR Institute of Complex Systems, Uos Sapienza, Piazzale Aldo Moro 2, 00185, Roma, Italy}
\alsoaffiliation{Department of Physics, Sapienza University of Rome, Piazzale Aldo Moro 2, 00185 Roma, Italy}
\author{Marco Laurati}
\email{marco.laurati@unifi.it}
\affiliation{Dipartimento di Chimica and CSGI, Università di Firenze, 50019 Sesto Fiorentino, Italy}

\title{Link between morphology, structure and interactions of composite microgels}

\begin{document}

\begin{abstract}
We combine small angle scattering experiments and simulations to investigate the internal structure and interactions of composite Poly(N-isopropylacrylamide)-Poly(ethylene glycol) (PNIPAM-PEG) microgels. At low temperatures the experimentally determined form factors and the simulated density profiles indicate a loose internal particle structure with an extended corona, that can be modeled as a star-like object. Increasing temperature across the volumetric phase transition, the form factor develops an inflection which, using simulations, is interpreted as arising from a configuration in which PEG chains are incorporated in the interior of the PNIPAM network. In this configuration a peculiar density profile characterized by two dense, separate regions is observed, at odds with configurations in which the PEG chains reside on the surface of the PNIPAM core. The conformation of the PEG chains have also profound effects on the interparticle interactions: While chains on the surface reduce the solvophobic attractions typically experienced by PNIPAM particles at high temperatures, PEG chains inside the PNIPAM network shift the onset of attractive interaction at even lower temperatures. Our results show that by tuning the morphology of the composite microgels we can qualitatively change both their structure and their mutual interactions, opening the way to explore new collective behaviors of these objects.
\end{abstract}

\section{Introduction}
Soft polymeric colloids display properties that are determined by the interplay between colloidal behavior and the features of the internal polymeric structure~\cite{lyon2012polymer}. The internal structure not only affects the single-particle properties, but also influences the particle-particle interactions~\cite{vlassopoulos2014tunable}. Within the family of polymeric soft colloids, microgels, in which the internal structure is made of a cross-linked polymer network with a typical core-corona architecture~\cite{stieger2004}, is a widely investigated system. The  polymer-colloid duality of this model system can be exploited to tackle fundamental physics problems, such as glass and jamming transition~\cite{saunders1999,yunker2014physics,brijitta2019,philippe2018glass} as well as to develop wide-ranging applications, including drug delivery systems~\cite{guan2011pnipam}, inks for 3D printing \cite{highley2019inks}, systems for CO$_2$ capture \cite{mengchen2014} and regenerative scaffolds~\cite{griffin2015accelerated}.

 The properties of microgels strongly depend on the nature of the constituent polymers, which determines how the microgels respond to the variation of, for instance, temperature~\cite{stieger2004}, pH~\cite{nigro2015dynamic}, or external fields~\cite{colla2018self}. Most studies have focused on thermoresponsive microgels made of poly(N-isopropylacrylamide) (PNIPAM) ~\cite{sierra2011swelling,Scheffold2020}, whose hallmark is the presence of the so-called  volume phase transition (VPT) in water at a characteristic temperature $T_{c}\sim$ 32 °C from a swollen state at low $T$ to a compact state at high $T$. This transition is linked to changes in the mechanical properties of the particles~\cite{Hashmi2009}: whereas the colloid is soft in the swollen state, it becomes stiffer above $T_{c}$, where also the presence of attractive interactions arises, leading ultimately to phase separation~\cite{Wu2003}. This description can be modified by adding ionic groups~\cite{colla2018self,del2020charge}, inducing non-spherical shapes during the synthesis process~\cite{crassous2012preparation,kruger2018catalyst,wolff2020soft}, or creating core/shell microgels~\cite{hendrickson2010design,suzuki2013multilayered,cors2017core}. Thus, microgels not only display a self-adaptive behavior to environmental changes, but can also be programmed to have specific response thanks to the precise knowledge of the topology of the network and to the different polymers used during the synthesis. In this way, the spectrum of microgel applications can be even enlarged, covering photonic devices~\cite{graf2002metallodielectric}, regenerative materials~\cite{rose2017nerve}, and biomaterial design~\cite{saxena2014microgel}, to name a few. 
 
Among the wide range of possible modifications, the inclusion of  poly(ethyleneglycol) (PEG) in the PNIPAM microgel network has the potential of increasing the bio-compatibility of the particles for drug delivery applications, and can also be used to tune the value of $T_c$ and the degree of deswelling associated with the VPT~\cite{APnipamPeg,EsSayed2019}. However, the microscopic origin of these phenomena is yet unclear. Since PEG can be considered unaffected by $T$ in the range where the VPT occurs, these effects must be related to the relative distributions of PEG and PNIPAM within the particles, to their interactions and to how these affect the internal structure as a function of $T$.     

To shed light on these mechanisms, in this work we investigate composite microgels of PNIPAM and PEG using a combination of experiments and numerical simulations. We characterize the effect of PEG chains on the morphology of the microgels across the VPT using Small-Angle Neutron (SANS) and X-ray (SAXS) scattering experiments. Numerical simulations are then used to rationalize the experimental findings by studying how the  distribution and conformations of PEG chains in the particles affect the PNIPAM network structure as a function of $T$. In particular, we show that while the presence of PEG inside of the composite microgel induces formation of two dense regions and a smaller particle size, the size of the particle increases when the PEG chains are distributed on the surface. A qualitative comparison with experimental results allows us to discriminate that for the composite microgels investigated in this study, the PEG chains are mostly located inside the PNIPAM network.
We also calculate the effective potential for each distribution, finding different behavior depending on the PEG chains distribution. On one hand, we find that addition of PEG chains on the surface of the microgels induce repulsive interactions even at temperatures above the VPT, thus effectively shielding the hydrophobic attraction between PNIPAM monomers.
On the other hand, we find that when the PEG chains are inside the microgels, attractive interactions arise even below the VPT, differently from standard PNIPAM microgels.

These results suggest the tuning the microgel morphology is a convenient way to tailor the structure and the interactions between the particles, which can be exploited in the future to vary the assembly and the rheology of these systems at high densities.

\section{Materials and Methods}

\subsection{In Vitro PNIPAM-PEG particles}
\subsubsection{Microgel Synthesis.} Composite PNIPAM-PEG microgel particles were synthesized following a ‘‘one pot’’ soapless emulsion polymerization method \cite{APnipamPeg}. 
All reagents were purchased from Sigma-Aldrich.  N-isopropylacrylamide ($\text{M}_{\text{n}}$ 113.16 g/mol) was purified by recrystallization in petroleum ether at 35 °C. The crosslinker ethylene glycol dimethacrylate (EGDMA), the initiator Ammonium Persulfate (APS) ($\text{M}_{\text{n}}$ 228.18 g/mol) and the Poly(ethyleneglycol) methyl ether methacrylate (PEG) ($\text{M}_{\text{n}}$ 950 g/mol) were used as purchased. 
The synthesis was carried out using a 1 L jacketed glass reactor (Syrris, model Atlas Potassium, Royston, U.K.) to improve temperature and stirring control. The particles were synthesized with a proportion in weight equal to 30\% PEG and 70\% PNIPAM. Initially 3.5 g of PNIPAM were dissolved in 40 ml of water and mixed with the EGDMA crosslinker (1 mol\% vs PNIPAM). The so-obtained solution was bubbled with nitrogen for 30 minutes to remove any dissolved oxygen while stirred at 350 rpm in a cold bath at 15 °C.  After 20 minutes, 1.5 g of PEG predissolved in 10 ml of water was added to the solution and the bubbling was maintained for 10 additional minutes.  The obtained mixture was added to 438 ml of preheated water (85 °C) and stirred at 350 rpm for 30 minutes. APS (2 wt\% vs PNIPAM) previously dissolved in 12 ml of water was added to initiate the reaction. The polymerization was carried for 45 minutes, after which the solution was placed in a cold bath to stop the polymerization process. The dispersion was purified via dialysis for 7 days and the microgel particles were recovered by freeze drying. The microgel particles were redispersed in deuterated water ($\text{D}_{2}\text{O}$) obtaining a diluted sample with concentration  $C=0.0010$ g/ml. Deuterated water was chosen to increase the contrast in neutron scattering measurements. Particle characterization obtained by Dynamic Light Scattering in previous work showed that the hydrodynamic radius $R_H\approx 166$~nm at small $T$ and that the volume phase transition (VPT) occurs at $T_c\approx36$ °C leading to $R_H\approx 90$~nm at high $T$~\cite{LaraPena2021}. This value of $T_c$ is sensibly larger than that usually found for standard PNIPAM microgels ($T_c\approx 32$ °C).

\subsubsection{Small-Angle Neutron Scattering (SANS) Measurements.} SANS measurements were performed at the NG7 SANS beamline (NCNR at NIST, Gaithersburg, USA) using three different configurations: {\it(i)} 1.33 m Sample-to-Detector Distance (SDD) and incident wavelength $\lambda$ = 6 \AA, {\it(ii)}  4 m SDD and $\lambda$ = 6 \AA, and {\it(iii)} 13.17 m SDD and $\lambda$ = 8.4 \AA. The combination of the three configurations gives a wave vector range 0.001 $\text{\AA}^{-1} < q < 0.4\text{ \AA}^{-1}$. The scattering length density of the different components of the samples were determined using the NIST scattering length density calculator ((\url{https://www.ncnr.nist.gov/resources/activation/})):  $\rho_{\text{PNIPAAm}} = 0.814 \times10^{-6} \text{ \AA}^{-2}$, $\rho_{\text{PEG}} = 0.599 \times10^{-6} \text{ \AA}^{-2}$, and $\rho_{\text{D}_{2}\text{O}} = 6.38 \times10^{-6} \text{ \AA}^{-2}$. Measurements were performed at 20 °C, 30 °C and 40 °C. 

\subsubsection{Small-Angle X-ray Scattering (SAXS) Measurements.} The SAXS experiments were performed at the Austrian SAXS Beamline at Elettra Sincrotrone Trieste (Trieste, Italy). X-Ray photons of energy $8$ keV, corresponding to a wavelength $\lambda$ = 0.154 nm, were used in the experiments. 
The $q$ range of the measurements was $0.035\text{ \AA}^{-1} < q < 0.8\text{ \AA}^{-1}$.
The sample was measured at 25 °C, 30 °C, 35 °C, 40 °C, 45 °C, 50 °C, 55 °C and 60 °C. Intensities from samples were corrected for the empty cell and solvent contributions.

\subsubsection{SANS Data Analysis.} 
The intensity profile or macroscopic cross section in a neutron scattering experiment on dispersions of colloidal particles is given by~\cite{BHiggins}
\begin{equation*}
I(q)=  \phi V(\Delta \rho)^{2} P(q)S(q)
\end{equation*}
where $\phi$ is the volume fraction occupied by the particles, $V$ the particle volume, $\Delta \rho=\rho_{1}-\rho_{2}$ the scattering length density difference between the particles ($\rho_1$) and the solvent ($\rho_2$), $P(q)$ the particle form factor, and $S(q)$ the structure factor. For dilute samples, as in this work,  $S(q)=1$ and the scattered intensity is proportional to the particle form factor $P(q)$.
Considering the small degree of cross-linking of the PNIPAM component we expect a very open particle structure. For this reason, following previous work on PNIPAM microgel particles cross-linked with PEG~\cite{AStarPolym} having a similar cross-linker density, we used the star polymer form factor model of Dozier and coworkers~\cite{AStarPolym2} to fit the experimental intensity profiles. The model consists of two terms:
\begin{equation}\label{eq:PStar}
P(q)= A_{1}e^{-\frac{1}{3}q^{2}R_{g}^{2}}  + A_{2}\frac{\sin(\mu\tan^{-1}(q\xi))}{q\xi (1+q^{2}\xi^{2})^{\mu/2}}
\end{equation}
The first term is a Guinier form factor, which yields a measure of the size of the particles through the radius of gyration $R_{g}$. The second term models the blob scattering of the star arms. The excluded volume correlation length or blob size $\xi$ is the characteristic length scale at which the granular polymer structure becomes relevant. 
The quantity $\mu$ is defined as $\mu=1/\nu-1$, where $\nu$ is the Flory exponent.
The amplitudes $A_{1}$ and $A_{2}$ weight the contributions of the total and internal terms of the model. Data modeling was performed with SasView~\cite{SAS}. 

\subsection{In Silico PNIPAM-PEG particles.} 
\subsubsection{Numerical Microgel Synthesis} Following previous well-established protocols~\cite{gnan2017silico}, microgels were numerically designed as fully-bonded, disordered networks resulting from the self-assembly of a binary mixture of limited-valence particles of diameter $\sigma_{m}$. Specifically, we used $N_{A}$ particles of species $A$ with two attractive patches to mimic monomers (N-isopropylacrylamide, NIPAM) and $N_{B}$ particles of species $B$ with four attractive patches to resemble crosslinkers (ethylene glycol dimethacrylate EGDMA). To reproduce the characteristic core-corona structure of the microgels, we also use an additional confining force acting on the crosslinkers only~\cite{ninarello2019modeling}. Once a fully-bonded network is  obtained, the polymer network structure is fixed by making bonds permanent. To do this, the patchy interactions are replaced by ones representative of polymeric systems, by using the Kremer-Grest bead-spring model~\cite{kremer1990dynamics}, where all particles interact via a Weeks-Chandler-Andersen (WCA) potential, defined as:
\begin{equation}
\label{eq:WCA}
V_{WCA}\left(r \right )=\left\{\begin{matrix}
4\epsilon\left[\left(\frac{\sigma_{m}}{r}\right)^{12}-\left(\frac{\sigma_{m}}{r}\right)^{6} \right] + \epsilon \mbox{ if $r\leq2^{1/6}\sigma_{m}$}& \\ 
0  \mbox{\hspace{32mm} otherwise} & 
\end{matrix}\right.
\end{equation}
\noindent where $\sigma_{m}$ is the unit of length and $\epsilon$ controls the energy scale. Additionally, bonded particles also interact via a FENE potential $V_{FENE}$:
\begin{equation}
\label{eq:FENE}
V_{FENE}\left(r \right )=-\epsilon k_{F}R_{0}^{2}ln\left(1-\left[\frac{r}{R_{0}\sigma_{m}}\right]^{2}\right) 
\end{equation}
\noindent where $k_{F}=15$ is the dimensionless spring constant and $R_{0}=1.5$ is the maximum extension of the bond. 

Finally, the thermoresponsive behavior of the PNIPAM microgels is captured by adding an effective attraction among monomers:
\begin{equation}
\label{eq:alpha}
V_{\alpha}\left(r \right )=\left\{\begin{matrix}
-\epsilon\alpha \mbox{\hspace{36mm} if $r\leq2^{1/6}\sigma_{m}$}& \\ 
\frac{1}{2}\alpha\epsilon\left[cos\left(\delta\left(\frac{r}{\sigma_{m}}\right)^{2}+\beta\right)-1\right] \mbox{\hspace{5mm} if $r<2^{1/6}\leq R_{0}\sigma_{m}$}& \\ 
0  \mbox{\hspace{36mm} otherwise} & 
\end{matrix}\right.
\end{equation}

\noindent with $\delta=\pi\left(2.25-2^{1/3}\right)^{-1}$ and $\beta=2\pi-2.25\delta$~\cite{soddemann2001generic}. The parameter $\alpha$ modulates the solvophobicity of the beads and plays the role of an effective temperature in the simulations~\cite{soddemann2001generic,verso2015simulation}: for $\alpha=0$ the effective attraction is not present and hence we can reproduce good solvent conditions. Previous works have shown that the VPT transition occurs at a critical value $\alpha_c \sim 0.65$~\cite{gnan2017silico,moreno2018computational}.

\subsubsection{\label{sim_pegchains}Addition of PEG chains.} Once the microgel is formed, we perform a second step in the numerical synthesis to incorporate the PEG chains into the polymeric network. To compare with the experimental observations, we consider three possible ways of distributing the chains within the swollen microgel, i.e., at $\alpha=0$: 
{\it (i)} one end of each chain is attached to a NIPAM monomer on the surface of the microgel, while the other end remains free; {\it (ii)} both ends of each chain are attached to PNIPAM monomers on the surface of the microgel; {\it (iii)} chains are inserted inside the microgel, allowing them to find accommodation for all beads via energy minimization. The system is then relaxed, and we allowed both ends of the chains to form links with PNIPAM monomers in the network. In this work, we refer to these three distributions conventionally  as {\it chains}, {\it loops} and {\it inside}, respectively.

The interaction between PNIPAM and PEG  also follows the Kremer-Grest bead spring model, although, for the PEG monomers, the effective attraction due to the thermoresponsivity is ignored because it is well-known that solvent quality effects become evident at a much higher temperature than for PNIPAM ones~\cite{chudoba2017temperature}, outside the effective temperature range investigated in this work.

\subsubsection{Determination of Structural Quantities.} The microgel radius of gyration is defined as:
\begin{equation}
\label{eq:Rg}
R_{g}=\left\langle \sqrt{\frac{1}{N}\sum_{i=1}^{N}\left(\overrightarrow{r}_{i}-\overrightarrow{r}_{cm}\right)^{2}} \right\rangle,
\end{equation}
\noindent where the brackets $\left\langle\cdot\right\rangle$ indicate ensemble averages, $\overrightarrow{r}_{i}$ is the position of the $i-$th monomer and $\overrightarrow{r}_{cm}$ is the microgel's center of mass.

The inner structure of the macromolecules was studied through the radial density profile: 
\begin{equation}
\label{eq:Rdp}
\rho\left(r\right)=\left\langle \frac{1}{N}\sum_{i=1}^{N}\delta\left( \left|\overrightarrow{r}_{i}-\overrightarrow{r}_{cm}\right|-r\right)\right\rangle\,,
\end{equation}

\noindent At the same time, the microgel form factor $P\left(q\right)$ was calculated from equilibrated trajectories using the following expression:
\begin{equation}
\label{eq:Pq}
P\left(q\right)=\left\langle \frac{1}{N}\sum_{i,j}^{N}\exp\left(-i\overrightarrow{q}\cdot\overrightarrow{r}_{ij}\right)\right\rangle\,,
\end{equation}
\noindent where $\overrightarrow{r}_{ij}$ is the distance between monomers $i$ and $j$. Here, angular brackets indicate an average over different configurations and orientations of the wavevector $\overrightarrow{q}$. In particular, we consider $300$ distinct directions randomly chosen on a sphere of radius $q$.

\subsubsection{Simulation parameters.} To match the experimental polydispersity, we simulated microgels with $N=5000$, $20000$, $31000$ and $42000$ beads, all of them at a crosslinker concentration $c=1\%$. Then, to fix the number of PEG chains $f$ and their contour length, defined as $L_{c}=N_{pol}b$, we run a set of simulations at $\alpha=0$ for the case where the chains are attached on the microgel by one end. Here, $N_{pol}$ corresponds to the number beads and $b$ is the minimum of the FENE interaction. In particular, we fixed $f$ and $N_{pol}^{Chains}$ so that $\rho\left(r\right)\rightarrow0$ happens at the same $L_{c}^{chains}/R_{g}^{M}$ for the four different microgels considered here, where $R_{g}^{M}$ refers to the radius of gyration for the microgel without PEG chains. On the other hand, for the {\it loops} case, we considered $L_{c}^{Loops}=2L_{c}^{chains}$ to ensure that $\rho\left(r\right)$ decays approximately at the same $r$ value than for the {\it chains} case. Finally, for the {\it inside} distribution, PEG chains were cut to guarantee that $L_{c}^{Inside}<R_{g}^{M}$. Table~\ref{tab:SimParameters} collects all parameters used for the different PEG distributions.

\begin{table}[ht]
\centering
\begin{tabular}{|c|c|c|c|c|}
\hline
$N_{microgel}$ & \textbf{$f$} & \textbf{$N_{pol}^{Chains}$} & \textbf{$N_{pol}^{Loops}$} & \textbf{$N_{pol}^{Inside}$} \\ \hline
5000                     & 90           & 10                          & 20                         & 4                           \\ \hline
20000                    & 90           & 70                          & 140                        & 28                          \\ \hline
31000                    & 90           & 90                          & 180                        & 36                          \\ \hline
42000                    & 90           & 120                         & 240                        & 48                          \\ \hline
\end{tabular}

\caption{Model parameters.}
\label{tab:SimParameters}
\end{table}

We perform Molecular Dynamics (MD) simulations of each composite microgel at different $\alpha$ values by using a Langevin thermostat to set the reduced temperature $T^{*} = k_{B}T/\epsilon=1$. All beads have unit mass $m$ and the integration time-step is $\delta t = 0.002 \sqrt{m\sigma_{m}^{2}/\epsilon}$. Once the simulations are properly equilibrated for each $\alpha$, we measure the observables explained above and we average them over microgels with different sizes in order to reproduce at best a similar polydispersity to the experimental sample. All simulations were made with LAMMPS~\cite{plimpton1995fast}.

\subsubsection{Assessment of the effective interaction potential.} The two-body effective potential between two composite microgels is evaluated by means of the umbrella sampling technique, where a series of independent configurations along a reaction coordinate are sampled by using a bias potential~\cite{likos2001effective,blaak2015accurate}. In this work, we consider the centers of mass distance of the macromolecules as the reaction coordinate, and the bias potential to be harmonic. Then, we evaluate the bias probability distribution $P_{b}\left(r,\Delta_{i}\right)$ of finding the macromolecules' centers of mass at distance $r$ given the equilibrium length of the spring $\mu_{i}$ from our simulations. Later, the contribution from the bias potential is removed, $P_{u}\left(r,\Delta_{i}\right)$ and subsequently unbiased probability distributions are merged into $P_{r}$ via a least-squares method. Thus, the potential of mean force is expressed as 
\begin{equation}
\label{eq:Veff}
V_{eff}\left(r\right)=-k_{B}T\left[P\left(r\right)\right]+C\,,
\end{equation}
\noindent being $C$ a constant that fixes the condition $V_{eff}\left(r\rightarrow  \infty \right)=0$.

\subsection{Calculation of the elastic moduli}
To calculate the microgel elastic moduli, we follow the approach that some of us recently developed~\cite{rovigatti2019connecting,camerin2020microgels}. From Mooney-Rivlin theory, the energy $U$ for a three-dimensional object due to thermal fluctuations can be written as a function of the invariants $J$ and $I$ of the strain tensor as

\begin{equation}
U\left(J,I_{1},I_{2}\right) = U_{0}+W\left(J\right)+W\left(I_{1}\right)+W\left(I_{2}\right)\,,    
\end{equation}
\noindent where $U_{0}$ is the energy of a reference configuration. This configuration is approximated to an average ellipsoid with  semi-axes $a_{1}$, $a_{2}$ and $a_{3}$, obtained by the gyration tensor built via the convex hull, defined on a set of points that encompasses all the particles of the composite microgel. These points are used for computing the gyration tensor, and hence, its diagonalization provide the three eigenvalues $\lambda_{1}$, $\lambda_{2}$ and $\lambda_{3}$, sorted from the largest to the smallest. Thus, the semi-axes are defined as $a_{i}=\sqrt{3\lambda_{i}}$. Likewise, $W$ depends on the potential of mean force, defined as

\begin{equation}
    W\left(X\right)=-k_{B}T lnP\left(X\right)+D_{X}
\end{equation}

\noindent where $X=J,I_{1},I_{2}$, $P\left(X\right)$ represents the respective probability distribution and $D_{X}$ is an arbitrary constant. These potentials can be fitted by using

\begin{equation}
    f\left(X;M_{X},X_{0},\gamma,C\right)=M_{X}\left(X-X_{0}\right)^{\gamma}+C
\end{equation}

\noindent with $\gamma=2$ for $X=J$ and $\gamma=1$ for $X=I$. In Fig. S3, we show the results for a composite microgel below and above the VPT. From $M_{J}$ and $M_{I}$ we  obtain the bulk modulus $K$ and the shear modulus $G$ as,

\begin{equation}
    K = \frac{2M_{J}}{V}
\end{equation}
\begin{equation}
    G = \frac{2M_{I}}{V}.
\end{equation}

\noindent The knowledge of these two elastic moduli allows us to calculate also the Young modulus $Y$ and the Poisson's ratio $\nu$ as

\begin{equation}
    Y = \frac{9KG}{3K+G} \\
\end{equation}
\begin{equation}
    \nu = \frac{3K-2G}{2\left(3K+G\right)}.
\end{equation}


\section{Results}
\subsection{Experiments}
\subsubsection{Experimental form factor obtained from SANS} 
\begin{figure}[ht]
	\centering
	\includegraphics[width=3.33in	]{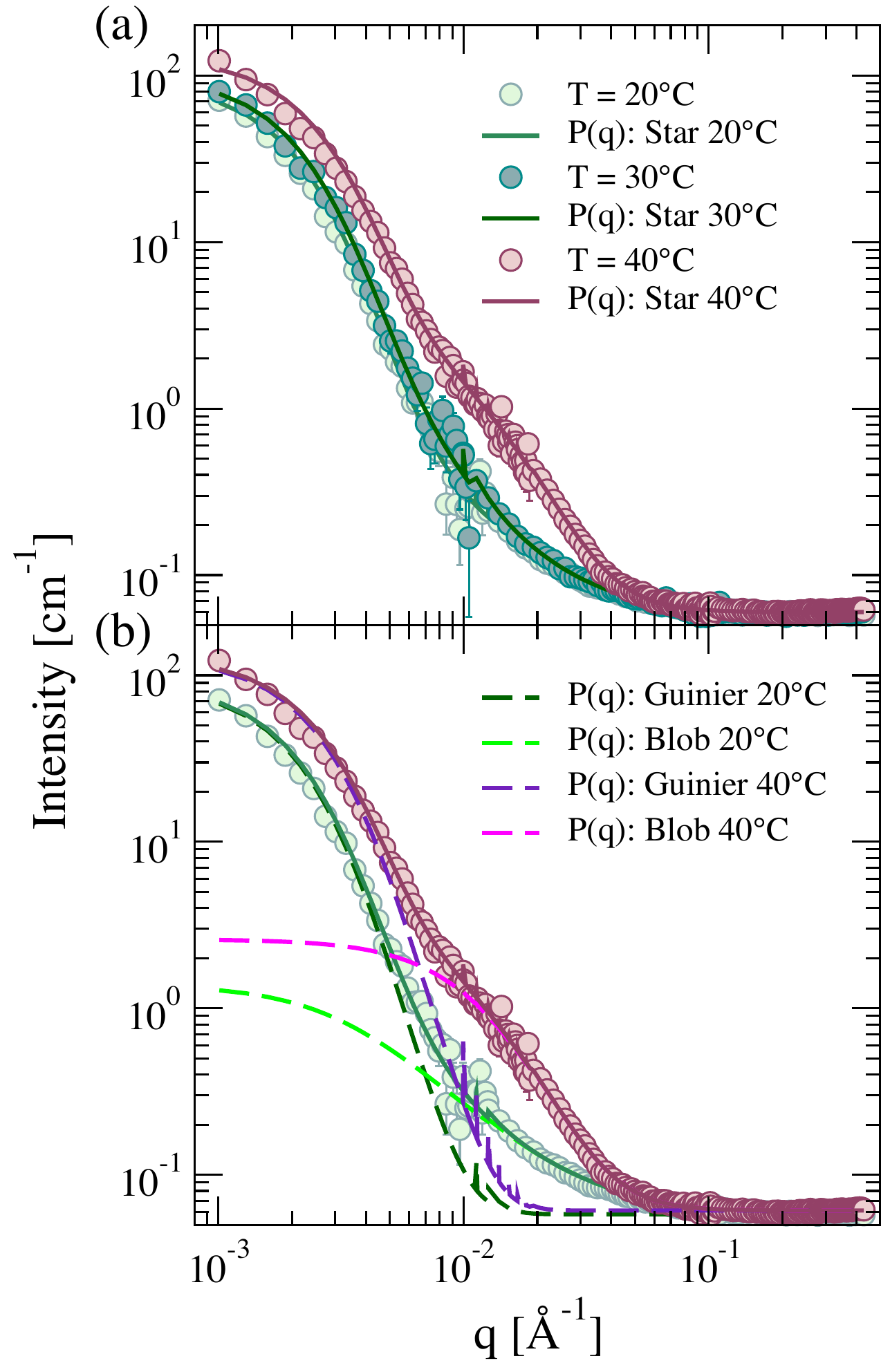}
	\caption{Intensity profiles and fits of sample with  $C=0.0010$ g/ml using the star polymer form factor model; (a) fits sample at 20 °C, 30 °C and 40 °C; (b) breakdown of the star polymer model in its two components.}
	\label{fig:PEG3CF}
\end{figure}

The intensity profiles measured at 20 °C, 30 °C and 40 °C are reported in Fig.\ref{fig:PEG3CF}.  Intensities for 20 °C and 30°C show a similar shape and the same order of magnitude. They present a smooth decrease as a function of $q$ typical of the form factor of diffuse soft structures, like star polymers and loosely cross-linked microgel particles~\cite{AStarPolym2}. Some larger fluctuations observed for 20 °C and 30 °C at $q$ values around $0.01 \text{ \AA}^{-1}$ are due to a non-perfect overlap between different experimental configurations of the sample-detector distance used to cover the reported range of $q$ values. At 40 °C, i.e. above the VPT temperature ($T_{c}\approx36$ °C), the profile shows a higher intensity and moves to larger $q$ values, indicating a reduction in the size of the dispersed particles. In addition an additional inflection point at intermediate $q$ values ($\sim 2\times10^{-2}\text{ \AA}^{-1}$) is present. 
The results of fitting the experimental SANS intensity profiles using the star polymer model of Eq.~\ref{eq:PStar} are shown as solid lines in Fig.~\ref{fig:PEG3CF}(a). 
Note that the fuzzy sphere model~\cite{stieger2004} typically used for microgels failed to properly describe the data, in agreement with a recent work where modifications on the topology of a PNIPAM microgel by the presence of an interpenetrating polymer network were also not well-described by the fuzzy sphere model~\cite{kozhunova2021microphase}.
At 20 °C and 30 °C the star polymer model nicely fits the experimental data at all $q$ values. There is also good agreement between the model and the measurements at 40 °C but with a slight overestimation of the model for the lowest $q$ values. 
The fitted parameters are listed in table~\ref{tab:StarParameters}. At 20 °C and 30 °C the radius of gyration is comparable, $R_g\approx 900\text{ \AA}$, while the blob size reduces from $\xi\approx350\text{ \AA}$ to $\xi\approx250\text{ \AA}$. The low value of the ratio $R_g/R_H\approx 0.56$ confirms the very open structure of our particles. At 40 °C $R_g\approx 740\text{ \AA}$, and the blob size reduces, $\xi\approx70\text{ \AA}$. The latter is mainly responsible for the appearance of the inflection at intermediate $q$ values, as shown in Fig.~\ref{fig:PEG3CF}b, where the contributions of the two terms in Eq.~\ref{eq:PStar} are shown separately for 20 °C and 40 °C.  The observed reduction of the radius of gyration and the blob size are in agreement with the expected worsening of the solvent  quality, which is confirmed by the increase of $\mu$ (table \ref{tab:StarParameters}). Note that at 40 °C the ratio $R_g/R_H\approx 0.63$ is consistent with a slightly more compact particle structure.

\begin{table}[ht]
	\centering
	\begin{tabular}{|c|c|c|c|} 
				\hline
				\multicolumn{4}{|c|}{$C=0.0010$ g/ml}\\
				\hline
				 & 20 °C &30 °C & 40 °C\\ 
				\hline \hline
				A1 (1/cm) & 90 $\pm$ 10.0  & 100 $\pm$ 7.5  & 130 $\pm$ 10.0\\  
				\hline
				A2 (1/cm) &  2.0 $\pm$  0.150 & 2.0 $\pm$  0.150  &  1.3 $\pm$  0.125\\  
				\hline  
				$\xi$ (\AA) &  320 $\pm$  10.0 &  250 $\pm$  12.5 &   70 $\pm$  3.0\\  
				\hline 
				$\mu$ &  0.66 $\pm$  0.10  & 1.0 $\pm$  0.05 &  1.95 $\pm$  0.07\\  
				\hline 
				$R_{g}$ (\AA) & 900 $\pm$  20.0 &  880 $\pm$  15.0 &  740 $\pm$  22.5\\  
				\hline 
				PD &  0.25 $\pm$  0.0075 & 0.25 $\pm$  0.0075  &   0.25 $\pm$  0.0150\\  
				\hline 
	\end{tabular}
	\caption{Star polymer model parameters obtained by fitting the experimental intensity profiles in Fig.~\ref{fig:PEG3CF}(a).}
	\label{tab:StarParameters}
\end{table}

\subsubsection{SAXS: Detailed deswelling behavior} 
\begin{figure}[ht]
	\centering
	\includegraphics[width=3.33in	]{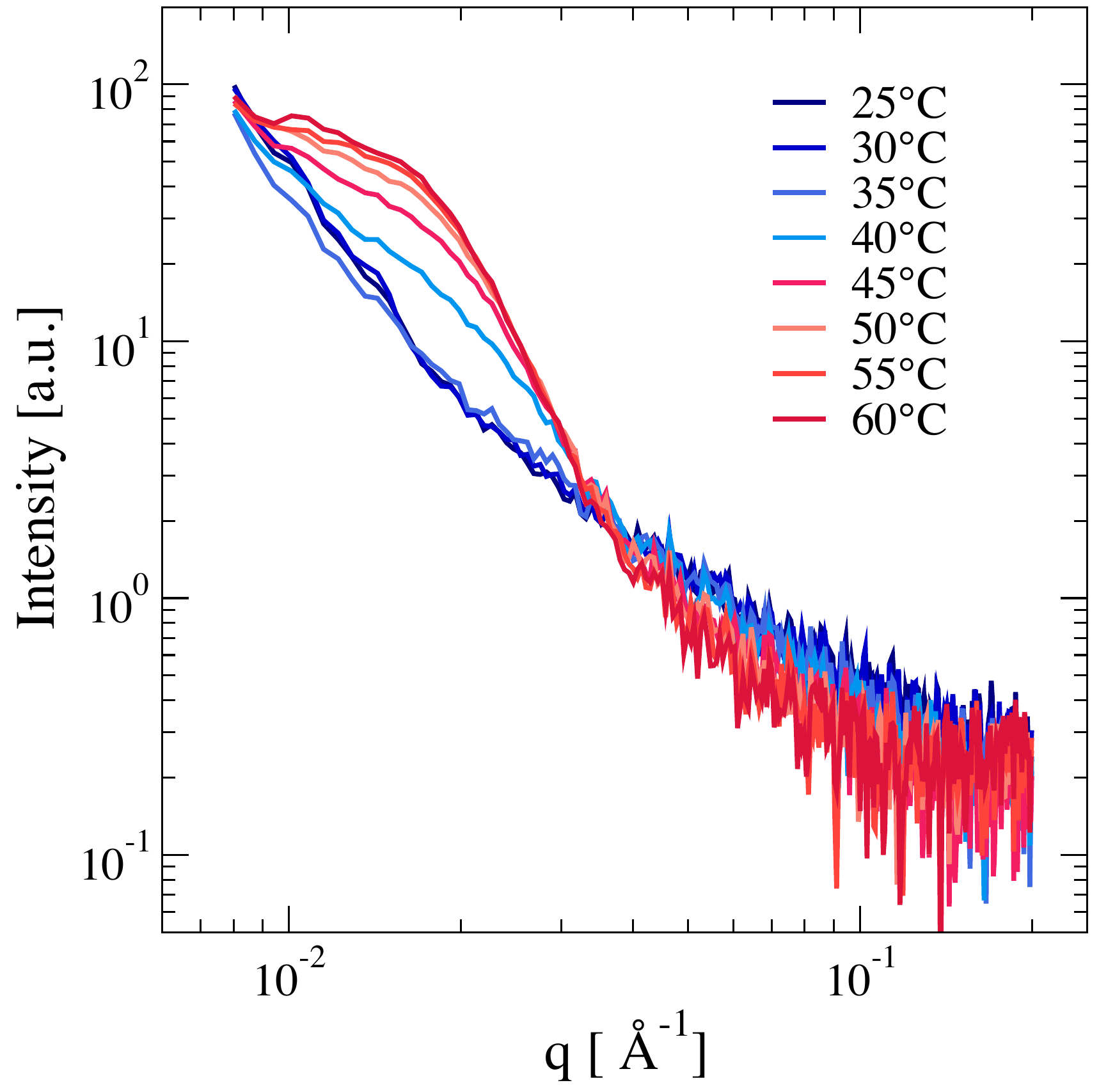}
	\caption{Intensity profiles obtained from SAXS at different temperatures, as indicated.}
	\label{fig:saxsall}
\end{figure}

Additional SAXS measurements of the intermediate $q$ range were performed to follow more in detail the deswelling behavior of the microgel particles. The intensity profiles obtained at 8 different temperatures are presented in Fig.~\ref{fig:saxsall}. At low $q$ values , near  $0.02 \text{ \AA}^{-1}$, profiles can be separated in three groups according to the intensity in this region:  {\it(i)} the first group includes profiles for 25 °C, 30 °C and 35 °C, which are very similar; {\it(ii)} the second group is formed by the profiles corresponding to 40 °C and 45 °C, which present a significantly larger scattering which increases with $T$, and the inflection already observed in the SANS data; finally {\it(iii)} the third group is formed by profiles corresponding to 50 °C, 55 °C and 60 °C, which present the largest scattering and the most pronounced inflection, but are now comparable with each other. 
Remembering that the critical temperature for these particles is $T_{c}\approx$ 36 °C, we may speculate that samples in the SAXS measurements were not fully equilibrated at the nominal temperature and that the real temperature of the measurements might be nearly 5 °C smaller than the nominal value. This assumption would be consistent with the idea that temperatures of group {\it(i)} lie below, temperatures of group {\it(ii)}  around, and temperatures of group {\it(iii)} above the VPT temperature. This is supported by the comparison of SAXS and SANS data presented in the Supplementary Material$^{\dag}$, which shows that SAXS data for 25 °C, 35 °C  and 45 °C overlap with SANS data for 20 °C, 30 °C and 40 °C, respectively. 
The SAXS data complement the SANS data showing that the deswelling transition is progressive and leads to the inflection of the form factor at intermediate $q$ values.
Next, we use simulations to assess the contribution of the PEG chains to the observed deswelling and blob shrinking.

\subsection{Simulations.}
The experimental form factors obtained by SANS and SAXS indicate a diffuse density profile, compatible with that of a star polymer, and a deswelling behavior which is characterized by a pronounced reduction of the blob size at higher temperatures, which together with the overall particle shrinking leads to the occurrence of an inflection of the form factor at intermediate $q$ values. However, due to the small contrast between PNIPAM and PEG, the analysis of the experimental data does not provide a clear indication of the distribution of PEG in the particles and how this affects the deswelling transition. To gain insight on these aspects we mimicked the experimental system in simulations resolved at the monomer scale.  
\begin{figure}[t]
	\centering
	\includegraphics[width=0.5\textwidth]{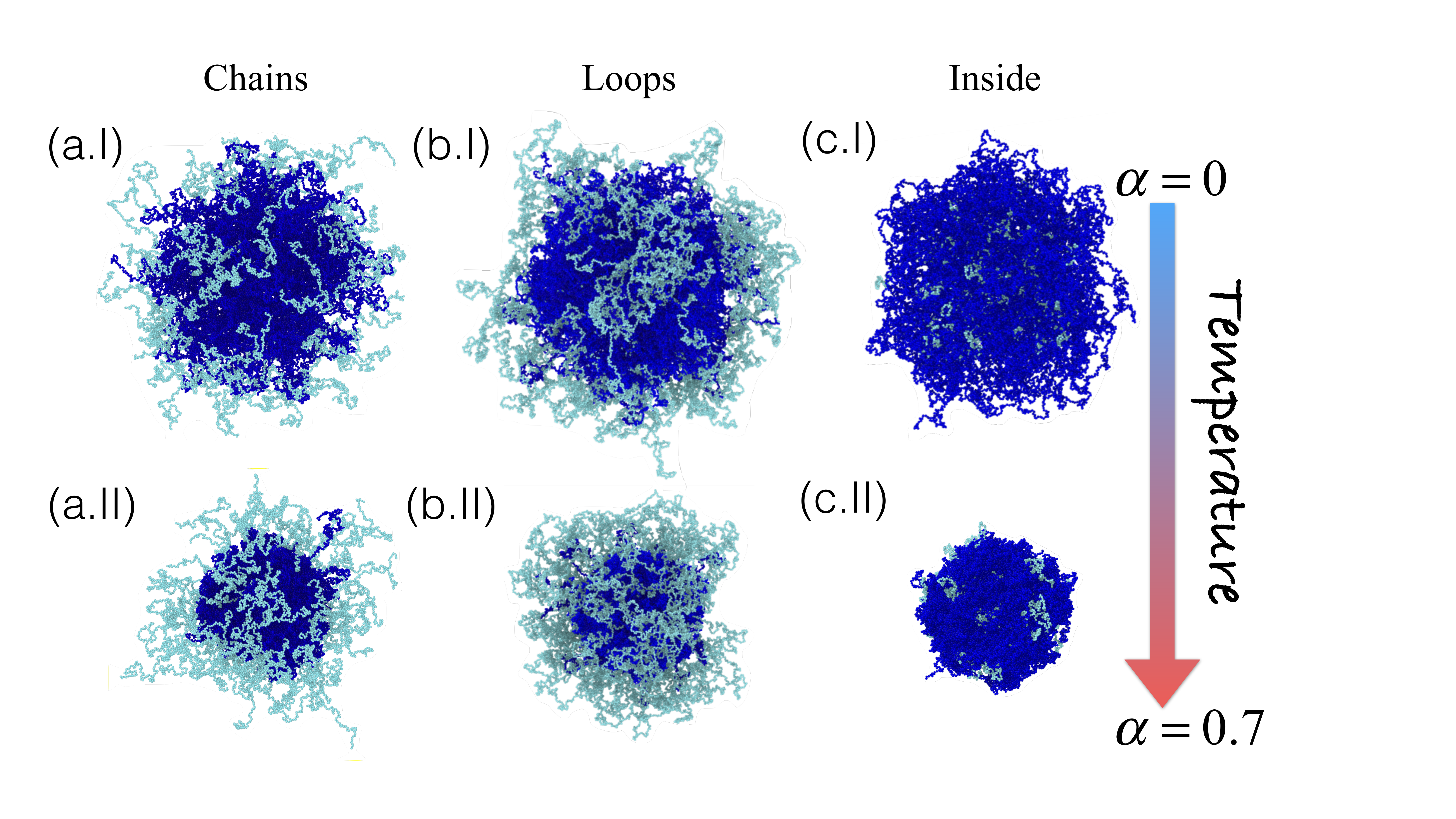}
	\caption{Snapshots to represent the structural change that the composite microgels undergo by increasing the temperatures when the PEG chains are distributed as (a) chains, (b) loops and (c) inside. Dark blue beads represent the PNIPAM microgel, whereas light blue monomers indicate the PEG polymer chains.}
	\label{fig:Pq}
\end{figure}

\subsubsection{Form factor: Effect of PEG distribution} 
\begin{figure}[!b]
	\includegraphics[width=3.33in	]{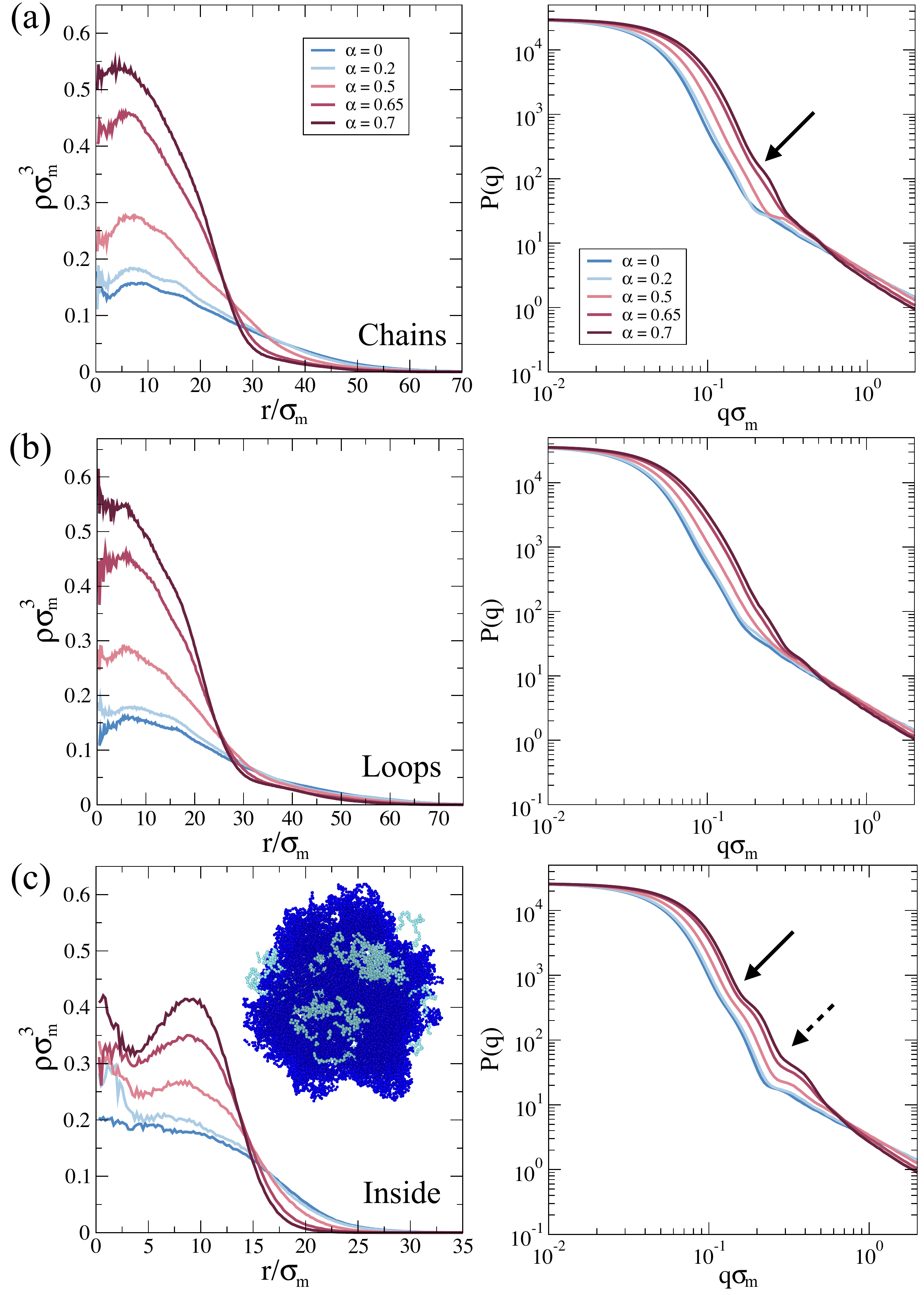}
	\caption{Numerical density profiles $\rho\left(r\right)$ and numerical form factors $P\left(q\right)$, averaged over the four different composite microgel, as a function of the solvophobic parameter $\alpha$, for (a) chains, (b) loops and (c) inside. Solid and dashed arrows highlight the two inflection points discussed in the text. The figure also reports a snapshot of a slice of the microgel for $\alpha=0.7$.}
	\label{fig:Density_Pq}
\end{figure}

We first study the effect of the PEG distribution on the form factors, looking for the one that leads to results qualitatively comparable to the experimental ones. As detailed in Sec. \emph{Addition of PEG chains}\ref{sim_pegchains}, three different PEG configurations, called \textit{chains} (PEG linear chains attached at one end to the surface of the PNIPAM particle), \textit{loops} (PEG linear chains attached at both ends to the PNIPAM particle) and \textit{inside} (PEG chains inserted inside of the PNIPAM network) are considered. Fig.~\ref{fig:Pq} shows snapshots of the resulting composite microgels as a function of the solvophobic parameter $\alpha$, which is equivalent to temperature in the experiments. As expected, the composite microgel shows a compacted core below the VPT due to the nature of the PNIPAM microgel by increasing $\alpha$. However, the overall structure is different depending on the PEG distribution. Indeed, in the cases where the PEG chains are protruding from the surface (\textit{chains} and \textit{loops}), we observe how an external layer is formed, resembling a core-shell particle. On the contrary, in the case where PEG is placed inside of the PNIPAM network (\textit{inside}), no clear difference with a pure PNIPAM microgel can be discerned, expect for a few PEG monomers that form sort of patches on the surface. To gain insight into the structural changes, we compute the numerical density profiles $\rho$ as a function of $\alpha$ for the three PEG distributions considered. This information is reported in Fig.~\ref{fig:Density_Pq}. For the \textit{chains} and \textit{loops} cases, shown in Fig.~\ref{fig:Density_Pq}(a) and (b), a dense core is well-localized at $r\sim10\sigma_{m}$ by increasing $\alpha$. Likewise, in the range $10\sigma_{m} < r \lesssim 30\sigma_{m}$, the typical corona observed in PNIPAM microgels is appreciated~\cite{ninarello2019modeling}, followed by a smooth decay corresponding to the PEG polymer. Instead, for the \textit{inside} case, we note that the core develops two peaks with $\alpha$, indicating that the collapse of the composite microgel is not homogeneous (see Fig.~\ref{fig:Density_Pq}(c) and its corresponding inset). The presence of the second peak is attributed to the fluctuations of the PEG chains inside the microgel pushing the PNIPAM outwards, and hence, creating a less dense intermediate region. Furthermore, we can notice how $\rho\rightarrow0$ at smaller $r$ compared to \textit{chains} and \textit{loops}. In addition to the presence of PEG on the surface, fluctuations of these monomers pull the corona, increasing in this way overall the microgel size.

Corresponding form factors $P\left(q\right)$ are shown in the right column of Fig.~\ref{fig:Density_Pq}. In agreement with the density profiles, we observe that the most remarkable structural difference generated by the PEG chains is when they are distributed inside the microgel. This result is confirmed by the existence of two inflections in the $P\left(q\right)$ of Fig.~\ref{fig:Density_Pq}c for the largest $\alpha$ value, analogous to the two peaks observed in the respective $\rho\left(r\right)$, indicating the presence of two dense regions. The inflections become increasingly pronounced with increasing solvophobicity, in qualitative agreement with the experimental results with increasing $T$. A smaller inflection develops with increasing solvophobicity also when the chains are instead distributed on the surface of PNIPAM and attached at one end (\textit{chains}). 

 Trying to link these results to the experimental findings, we notice that the first inflection (present for the \textit{loop} and \textit{inside} cases) arises at $q\sigma_{m} \sim 0.1$ while the second one (only present in the \textit{inside} case) for $q\sigma_{m} \sim 0.3$. In order to compare these numbers with experimental units, we need an estimate for $\sigma_{m}$, that is the diameter of the monomer in the simulation. Using the experimental value of $R_g$, we get $\sigma \approx 5$ nm, so that the two inflections should be located around $3\times 10^{-3}$ \AA$^{-1}$ and $ 10^{-2}$ \AA$^{-1}$, respectively. While the first one is not present in the experimental data, probably due to the large polydispersity of the microgels, the second one is evident in both SAXS and SANS data. This suggests that the simulated microgel with inside chains is the closest topology to the experimental system.

\subsubsection{Swelling behaviour.} 
\begin{figure}[!t]
	\centering
	\includegraphics[width=3.33in]{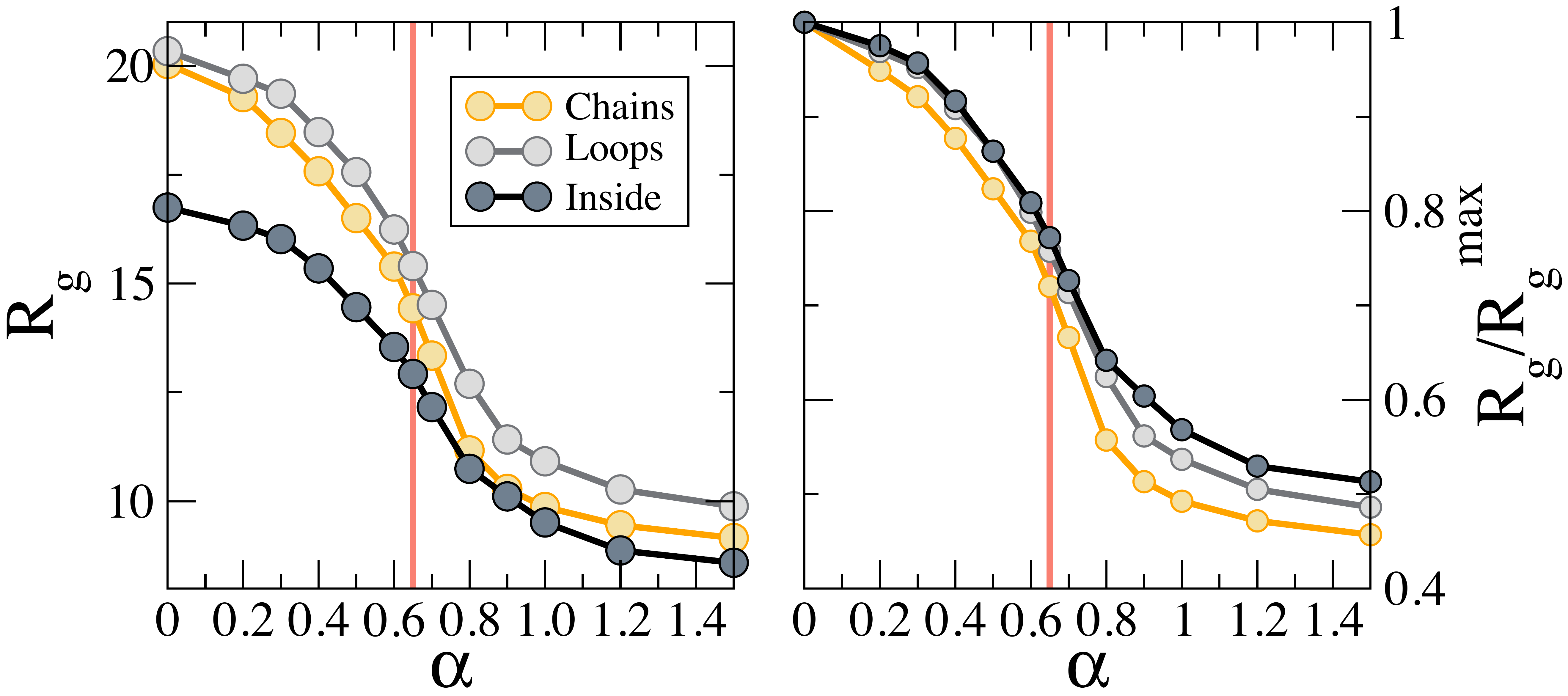}
	\caption{Swelling curves. (a) Radius of gyration $R_{g}$ as a function of the solvophobic parameter $\alpha$ for microgels with $N=5000$, for the three different distributions of the PEG polymer chains analyzed in this work. (b) Normalized radius of gyration by $R_{g}^{max}$, to highlight the evolution of particle size as a function of how the chains are distributed. Vertical red line is placed at $\alpha\sim0.65$, indicating the VPT for a PNIPAM microgel.}
	\label{fig:Rg}
\end{figure}

In this section we focus on different swelling stages of the composite microgels upon increasing the solvophobicity parameter for the three different distributions of the PEG chains considered. In Fig.~\ref{fig:Rg}(a), we represent the swelling curves of a microgel with $N=5000$ beads. In agreement with the density profiles discussed in Fig.~\ref{fig:Density_Pq}, we observe that the composite microgel size is larger when the chains are on the polymer network's surface. 
In particular, for the case where both ends are connected to form loops, the composite microgel always acquires a slightly larger size than the case of chains. This situation, that is enhanced by increasing $\alpha$, resembles what already observed in charged co-polymerized microgels~\cite{del2020charge}, where charges tend to swell the network. On the other hand, when the distribution of chains is in the interior, we see that the composite microgel size is the smallest. This is due both to the short chains used in the simulations (see Table~\ref{tab:SimParameters}) and to the fact that in this configuration the microgel remains overall more compact.

At the same time, we focus on the volume phase transition (VPT), which has been documented to happen at $\alpha\sim0.65$ for a pure PNIPAM microgel~\cite{ninarello2019modeling}. Thus, in Fig.~\ref{fig:Rg}(b) we normalize the radius of gyration by its value at $\alpha=0$, that is, the maximally swollen size, $R_{g}^{max}$. We can appreciate that the VPT transition seems to occur at slightly larger values of $\alpha$ compared to a pure PNIPAM microgel. In particular, particles with polymer chains forming loops on the surface of the microgel and with polymer chains distributed inside, show a greater deviation from the VPT observed for pure PNIPAM microgels. While fluctuations of polymer chains on the surface, as in the \textit{loop} case, seem to justify this behavior, the shift in the \textit{inside} case can be attributed to the presence of two dense regions and a depleted region in the composite microgel (Fig.~\ref{fig:Density_Pq}). The presence of a depleted region prevents PNIPAM monomers from concentrating in the center of the microgel, thus diverting the collapsed regime volume, as we can observe at $\alpha=1.5$. 

\subsubsection{Effective potential interactions.}
\begin{figure}[t]
	\centering
	\includegraphics[width=3.33in	]{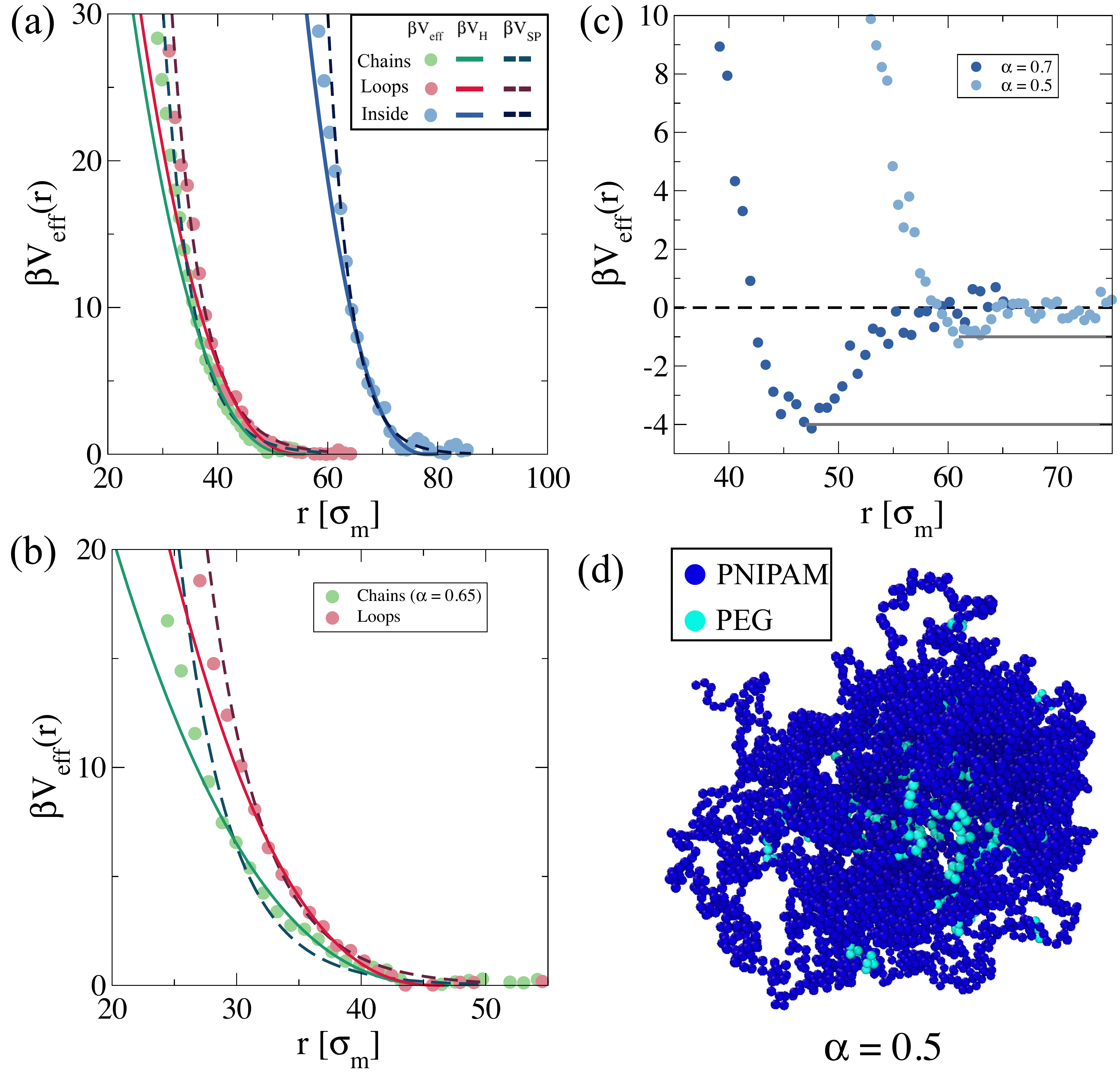}
	\caption{Effective interaction potential $\beta  V_{eff}$ computed by the umbrella sampling technique and fitted by the Hertzian model (Eq.~\ref{eq:Hertfit}) and the star polymer interaction (Eq.~\ref{eq:Starfit}) at (a) $\alpha=0$ for the three different distribution of the polymer PEG chains, (b) for the chains and loops cases at $\alpha=0.65$, and (c) for the inside case at $\alpha=0.5$ and $\alpha=0.7$ (d) Snapshot of the composite microgel with PEG chains inside at $\alpha=0.5$, showing it is rather non-spherical in shape due to the fluctuations of the chains inside the PNIPAM microgel.}
	\label{fig:pot}
\end{figure}

We study the effects of PEG chain distribution on the interaction of two PNIPAM-PEG microgels by computing the effective potential interaction $\beta V_{eff}\left(r\right)$. Results are shown in Fig.~\ref{fig:pot} for the three examined cases. 
The numerical results for the effective interaction at $\alpha=0$ are also compared to expected theoretical expressions. In particular, we consider the Hertzian model, usually employed for microgels~\cite{bergman2018new,rovigatti2019connecting,scotti2019deswelling}, which reads as
\begin{equation}
    \beta V_{H}\left(r\right)=U\left(1-\frac{r}{\sigma_{eff}^{H}}\right)^{5/2}\theta\left({\sigma_{eff}^{H}-r}\right)\,,
\label{eq:Hertfit}
\end{equation}
\noindent where $U$ is the Hertzian strength related to the elastic energy cost of particle deformation when they are pushed together and $\sigma_{eff}^{H}$ is an effective particle diameter beyond which the interaction vanishes, indicated by the Heaviside step function $\theta$. Since we earlier noticed that the experimental SANS intensity profiles cannot be fitted by the standard fuzzy sphere model, but we had to resort to a star polymer model, we also test whether a star polymer effective potential is applicable to describe the interactions of the  particles under investigation. For this reason, we also compare the numerical $V_{eff}$ to the star polymer potential developed by Likos and coworkers~\cite{likos1998star}, defined as,
\begin{equation}
    \beta V_{SP}\left(r\right)=\frac{5}{18}f^{3/2}\left\{\begin{matrix}
-ln\left(\frac{r}{\sigma_{eff}^{SP}} \right ) + \frac{1}{1+\sqrt f/2} & \textnormal{for  } r \leq \sigma_{eff}^{SP}\\ 
\frac{\sigma_{eff}^{SP}/r}{1+\sqrt f/2}\left[-\frac{\sqrt f\left(r-\sigma_{eff}^{SP}\right )}{2\sigma_{eff}^{SP}} \right] & \textnormal{for  } r \geq \sigma_{eff}^{SP}
\end{matrix}\right.,
\label{eq:Starfit}
\end{equation}
\noindent where $f$ is the number of arms, called functionality, and $\sigma_{eff}^{SP}$ represents the characteristic length of a star polymer.

Fig.~\ref{fig:pot}(a) shows $\beta V_{eff}\left(r\right)$ at $\alpha=0$ and the fits corresponding to the Hertzian and star polymer potentials. We find that the interaction is repulsive for all PEG distributions, as expected.
In particular, for the chains and the loop case the star polymer potential is found to properly capture the numerical results in the whole investigated range of $k_{B}T$ using $f=90$, the value used to simulate the form factors for comparison with the experimental results, and leaving $\sigma_{eff}^{SP}$ as the only fit parameter.  We estimate $\sigma_{eff}^{SP} = 28.5\sigma_{m}$ and $30\sigma_{m}$ for the \textit{chains} and  \textit{loops} configurations, respectively. On the other hand, when we fit the data with the Hertzian model using both $\sigma_{eff}^{H}$ and the interactions strength $U$ as fit parameters, deviations already appear for $\beta V_{eff} \gtrsim 10 k_{B}T$ in analogy to previous observation for standard microgels~\cite{rovigatti2019connecting}. In this case, the interaction lengths are $\sigma_{eff}^{H} = 54\sigma_{m}$ and $52.73\sigma_{m}$ for chains and loops, respectively. The remarkable difference in the size of the composite microgel between the two models is expected~\cite{parisi2020static}. Indeed, while $\sigma_{eff}^{SP}$ is well within the region of the outer blob chains, $\sigma_{eff}^{H}$ resembles the hydrodynamic radius $R_{H}$. 

We now turn our attention to the case in which the PEG chains are distributed within the PNIPAM microgel, the \textit{inside} case. From Fig.~\ref{fig:pot}(a) we note that the Hertzian model again starts to fail for $\beta V_{eff} \sim 10 k_{B}T$, with $\sigma_{eff}^{\left(H\right)}=78.48\sigma_{m}$. On the other hand, the star polymer potential describes the interaction up to $\beta V_{eff} \sim 20 k_{B}T$ when leaving $f$ and $\sigma_{eff}^{SP}$ both as fit parameters. We find $f=550$ and $\sigma_{eff}^{SP}=51.1\sigma_{m}$. Such a large value of the functionality may reflect the difference in the internal density of the particles evidenced in Fig.~\ref{fig:Density_Pq}(c), i.e. a heterogeneous internal structure that can be described as a high- functionality star polymer, in which the functionality arises from a mixture of the PNIPAM and PEG chains. This is markedly different from previous results on PNIPAM microgels, that even in the case of a low degree of cross-linking could be described with a fuzzy sphere model

Next, we increase $\alpha$ to study the effective potential close to the VPT. Results for \textit{chains} and \textit{loops} are shown in Fig.~\ref{fig:pot}(b). Previous computational studies of PNIPAM microgels have explored the two-body effective potential up to $\alpha=0.5$, i.e. below the VPT, in order to avoid the overall attraction between the two microgels~\cite{rovigatti2019connecting}. However, PEG monomers distributed on the surface turn out to effectively shield this attraction, inducing a completely repulsive interaction even above the VPT temperature. 
We can thus repeat at high temperatures the analysis made for $\alpha=0$ and compare the effective potentials obtained from simulations to the Hertzian and star effective potentials.  For \textit{chains} we find that the star polymer model (with $f=90$) is not particularly accurate in the description of the effective potential: this is mainly visible at intermediate distances where $V_{SP}$ appreciably underestimates $V_{eff}$. Instead, the Hertzian model seems to capture better the interaction in this range, but, as usual, only up to $\sim 10 k_{B}T$. On the other hand, $\beta V_{eff}$ for \textit{loops} seems to be well described by both models. 
The situation for the \textit{inside} case at high temperatures is instead completely different, as reported in Fig.~\ref{fig:pot}(c). Indeed, we find that for $\alpha=0.5$ an attraction is already present, with a minimum $\approx -1k_{B}T$, suggesting the existence of microgel-microgel attractive interactions even well below the VPT for this system. When we increase further the temperature ($\alpha=0.7$),  the attraction becomes stronger, reaching $\approx -4k_{B}T$, as shown in Fig.~\ref{fig:pot}(c). We highlight that the umbrella sampling technique is not very efficient for studying systems with too strong attractive interactions. Indeed, this technique is based on quantifying the transition probability between different states. Thus, the presence of strong bonds hinders an efficient exploration of the phase space, introducing a bias on the probability transitions. For this reason results at $\alpha = 0.7$ should be taken with caution. However, for $\alpha=0.5$ the involved energies are still not too large and data are reliable. 

A microscopic explanation of the onset of the attraction already below the VPT could be related to the fact that in this configuration the increase of $T$ induces a higher monomer density close to the external surface  of the microgel, resulting in a stronger effect of the solvophobic attraction when two particles are sufficiently close. 

This idea is supported by the snapshot shown in Fig.~\ref{fig:pot}(d), where it can be seen long PNIPAM chains are present on the surface. Notice also that the particle shape indicate a certain degree of anisotropy.

\begin{table}[ht]
\centering
\begin{tabular}{|c|c|c|c|c|}
\cline{1-5}
$\alpha=0$ & $K \left(\times10^{-3}\right)$ & $G \left(\times10^{-3}\right)$ & $Y \left(\times10^{-3}\right)$ & $\nu$ \\ \hline
\multicolumn{1}{|c|}{Chains} & $1.92$ & $1.56$  & $3.7$  & $0.18$            \\ \hline
\multicolumn{1}{|c|}{Loops}  & $2.6$  & $1.41$  & $3.7$  & $0.27$            \\ \hline
\multicolumn{1}{|c|}{Inside} & $0.42$  & $0.56$  & $1.16$  & $0.04$           \\ \hline
\end{tabular}
\caption{Elastic moduli for the composite microgel changing the polymer PEG chain distribution in units $k_{B}T/\sigma_{m}^{3}$. Whereas the error for $K$ and $G$ is $\lesssim 10\%$, for $Y$ and $\nu$ is $\sim 20\%$.}
\label{table:alpha0}
\end{table}

\begin{table}[ht]
\centering
\begin{tabular}{c|c|c|c|c|}
\cline{2-5}
& $K \left(\times10^{-3}\right)$ & $G \left(\times10^{-3}\right)$ & $Y \left(\times10^{-3}\right)$ & $\nu$ \\ \hline
\multicolumn{1}{|c|}{Chains ($\alpha=0.65$)} & $3.98$ & $3$  & $7.2$  & $0.2$            \\ \hline
\multicolumn{1}{|c|}{Loops ($\alpha=0.65$)}  & $4.72$  & $3.21$  & $7.9$  & $0.22$        \\ \hline
\multicolumn{1}{|c|}{Inside ($\alpha=0.7$)} & $4.07$  & $2.43$  & $6.08$  & $0.25$        \\ \hline
\end{tabular}
\caption{Elastic moduli for the composite microgel changing the polymer PEG chain distribution, in units $k_{B}T/\sigma_{m}^{3}$. Whereas the error for $K$ and $G$ is $\lesssim 10\%$, for $Y$ and $\nu$ is $\sim 20\%$.}
\label{table:alpha0.7}
\end{table}

Finally, we studied how the presence of PEG affects the elastic properties of the microgels. For this we followed the method recently proposed by some of us~\cite{rovigatti2019connecting, camerin2020microgels}. The resulting moduli are reported in Table~\ref{table:alpha0} and Table~\ref{table:alpha0.7} for $\alpha=0$ and $\alpha=0.7$, respectively. 
In particular, we find that the microgels with inside chains are sensibly less stiff than the other two cases: indeed, the bulk modulus $K$, the shear modulus $G$ and the Young modulus $Y$ are all smaller by roughly a factor of two with respect to the chains and loops. We notice that the latter (\textit{loops} case) seems to be slightly stiffer than the former (\textit{chains} case). These results are in line with expectations and are valid at all temperatures, for which the moduli are found to increase (in the investigated $\alpha$-range). Results for $\nu$ are much noisier, resulting from indirect estimate through $K$ and $G$ and we find that $\nu$ is roughly constant with temperature at around 0.2 for chains and loops, while it is much smaller for the inside case at $\alpha=0$. These considerations are in agreement with the working hypothesis of this study, i.e. that the microgels with inside PEG chains display a more heterogeneous internal structure characterized by the presence of a less dense region. 
Having estimated the elastic moduli, we can now perform a more stringent test of the Hertzian model, since the Hertzian stength $U$ in Eq.~\ref{eq:Hertfit} is related to the elastic properties as $U=\frac{2Y{\sigma_{eff}^{H}}^{3}}{15\left(1-\nu^{2}\right)}$. In this way, $\sigma_{eff}^{H}$ becomes the only fitting parameter and the resulting fits are discussed in the Supplementary Information (Fig. S3.) We confirm that the Hertzian fit with the true elastic constants works up to a few $k_B T$ for chains and loops, where the extracted $\sigma_{eff}^{H}$ is consistent with expectations. On the other hand, the fits is very poor for the inside case, where $\sigma_{eff}^{H}$
becomes much larger (contrary to the results for example of Fig.~\ref{fig:Rg}), again supporting the fact that the description with a star polymer potential works better in this case.

\section{Summary and Conclusions}

Combining scattering experiments and simulations we investigated the morphology and interactions of composite PNIPAM-PEG particles across the VPT transition. Experimental form factors $P\left(q\right)$ obtained from SANS and SAXS showed a collapse of the composite microgel by increasing $T$ beyond $T_{c}$. This behavior, also observed for PNIPAM microgels, is accompanied by unusual structural features. In particular, the fuzzy sphere model, which typically provides a good description of the form factor of PNIPAM microgels, fails to describe the PNIPAM-PEG composites. Instead, a star polymer model is able to capture the shape of $P\left(q\right)$. The observed star-like density profile can be the result of the lower degree of cross-linking with respect to PNIPAM microgels investigated in previous studies\cite{stieger2004,gasser2014}, leading to a particularly extended and diffuse corona. In addition, the deswelling associated with the increase of $T$ is associated with the appearance of an inflection of the experimental form factor at $q\sim10^{-2} \text{ \AA}$, which is not observed for simple PNIPAM microgels.\\ 
We used simulations to clarify how these unusual structural features are related to the conformations and relative spatial distribution of the PNIPAM and PEG chains. Simulations showed that when the PEG chains are distributed on the external surface of the PNIPAM network, either linked to the surface at one (\textit{chains} case) or both (\textit{loops} case) ends, the density profile is characterized by a denser core and a diffuse corona, with the corona progressively shrinking with increasing $T$. When the chains are only linked at one extreme, the decay of the density profile beyond the core becomes particularly sharp for high $T$, leading to the formation of a small inflection in the form factor $P\left(q\right)$ at intermediate $q$ values ($q\sigma_{m} \sim 0.1$). When the PEG chains are instead located inside the PNIPAM network (\textit{inside}), the structural evolution is significantly different: the density profile develops two denser regions with increasing $T$, with two separate decays that result in two inflections of $P\left(q\right)$, one at $q\sigma_{m} \sim 0.1$ as in the previous case, and the second at a larger value ($q\sigma_{m} \sim 0.3$). The $q$ value corresponding to the second inflection roughly corresponds to that of the experimental form factor. This indicates that the \textit{inside} configuration might reflect the morphology of the experimental system. We should consider however that the experimental system might present different coexisting configurations, and thus correspond to a combination of the \textit{chains} and the \textit{inside} configuration. Indeed, the synthetic method used in this work resembles a precipitation polymerization, since at the beginning NIPAM monomers are water soluble and PEG-methacrylate chains also, so the initially distribution of NIPAM and PEG is random; however when the polyNIPAAm chains start to grow the high polymerization temperature in water results in the tendency of polyNIPAAM chains to precipitate; however this is prevented by the PEG-methacrylate stabilization, leading to the progressive building of the core. Further polymerization occurs in a core shell manner, resulting in a soft core including some PEG-chains inside and additional, newly formed PEG chains attached to the surface of the microgel. The fact that the \textit{inside} configuration significantly differs from common PNIPAM internal structures is also confirmed by the investigation of the elastic properties of microgels, which evidence lower moduli (bulk, shear and Young modulus) compatible with a less denser/compact particle.\\
The simulations additionally show a very good agreement with previous experimental data\cite{LaraPena2021} concerning the reduction of the particle radius with increasing $T$ ($\alpha$), which is about  $50$\% in both cases. Note that this reduction would be significantly larger in pure PNIPAM particles synthesized with the same method and cross-linking density\cite{APnipamPeg}. The experimental data in\cite{LaraPena2021} correspond to measurements of the hydrodynamic radius using dynamic light scattering. The value of $R_g$ obtained through the modeling of the SANS experimental data presented here shows a smaller relative reduction, suggesting that the $R_g$ value estimates the size of the denser core of the particles.
Finally, simulations confirm experimental results\cite{APnipamPeg} indicating a higher value of $T_c$ for the PNIPAM-PEG particles.\\
The presence of PEG chains and their configurations also alter the interaction potential between two composite microgels. On one hand, at low $T$, the microgel stiffness is larger in the presence of PEG chains inside the microgel: This is the results of the presence of a softer PEG shell in the \textit{chains} and \textit{loops} cases. On the other hand, when $T$ increases the same PEG shell  inhibits the attraction induced by the solvophobic character of the PNIPAM polymer network. Instead, for the \textit{inside} case, the development of such an attraction occurs even below the VPT. This is attributed to the change in the particle density profile observed for this case, which results in a higher density of PNIPAM chains close to the surface. These results indicate that the PEG/PNIPAM composition could be varied to trigger the onset of attractive interactions at well-defined temperatures, even in swollen-like conditions, which is a fascinating possibility in the soft particle realm, very promising to observe new phase and rheological behavior~\cite{vlassopoulos2014tunable}.

\begin{acknowledgement}

R. R.-B., M. L.-P. and M.L. acknowledge support from the project ‘‘A1-S-9098” funded by Conacyt within the call ‘‘Convocatoria de Investigación Básica 2017-2018” and from ‘‘Consorzio per lo Sviluppo dei Sistemi a Grande Interfase” (CSGI). J.R.-F., F.C. and E.Z. acknowledge support from the European Research Council (ERC Consolidator Grant 681597, MIMIC). R. R.-B. and E.Z. acknowledge funding from H2020 Marie Curie Actions of the European Commission (ITN SUPERCOL, Grant Agreement 675179).

\end{acknowledgement}

\begin{suppinfo}

SANS and SAXS intensity profiles comparison, fuzzy sphere model form factor fit, and simulation details of the Hertz modulus are available in the supplementary material document.

\end{suppinfo}

\bibliography{biblio}

\end{document}


\maketitle

\section{Comparison of SANS and SAXS results}
\begin{figure}[ht]
	\centering
	\includegraphics[width=0.75\textwidth]{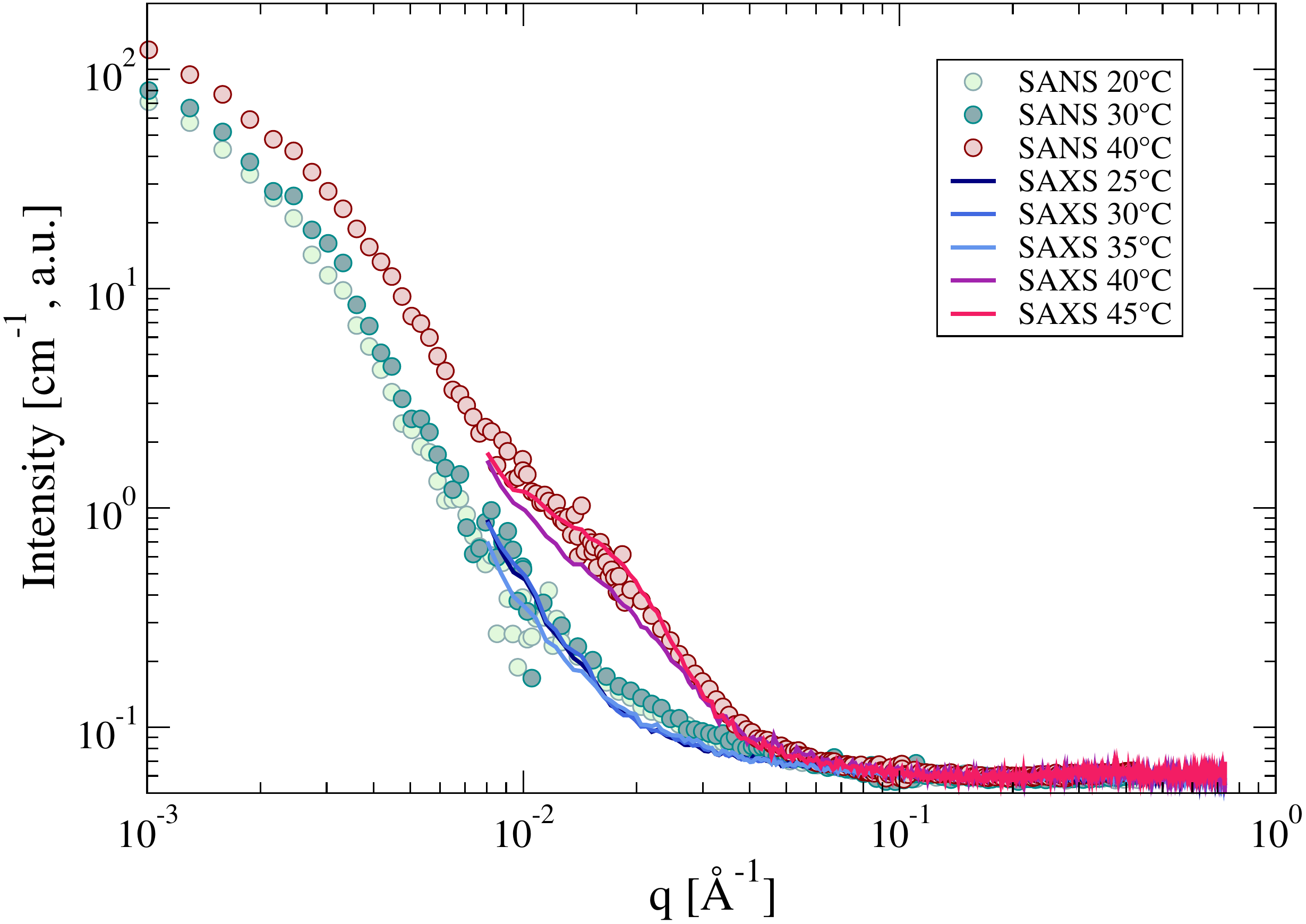}
	\caption{SANS and SAXS intensity profiles. The SAXS data has been scaled with different parameters for the (25°C, 30°C, 35°C) and (40°C, 45°C) sets.
	SANS intensity in $\text{cm}^{-1}$ units, SAXS intensity in arbitrary units.}
	\label{fig:NX2030}
\end{figure}

Fig.~\ref{fig:NX2030} shows a comparison of the SANS and SAXS measurements after the SAXS data sets have been scaled in the {\it y}-axis plus added the SANS background.
After applying scaling and shifting, excellent qualitative agreement between the profiles obtained with the two techniques is observed in the overlapping $q$ range, when SAXS data are compared with SANS data at a temperature about 5 °C lower. This supports the speculation proposed in the manuscript that the effective sample temperature of the SAXS measurements was about 5 °C lower than the nominal temperature. The SAXS data complement the SANS data showing that the deswelling transition is progressive and leads to the inflection of the form factor at intermediate $q$ values.

\FloatBarrier
\section{Fuzzy Sphere}

\begin{figure}[ht]
	\centering
	\includegraphics[width=0.75\textwidth ]{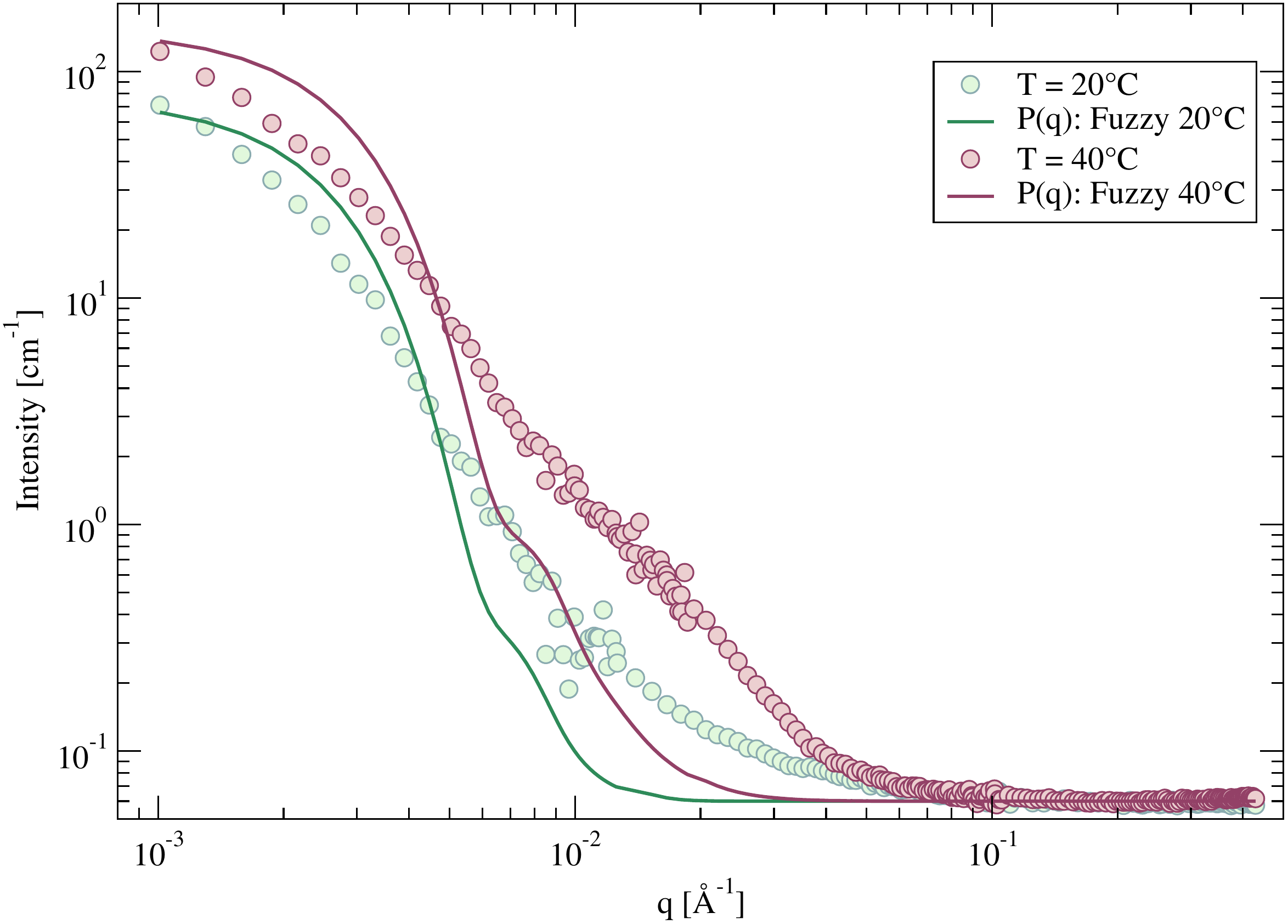}
	\caption{Intensity profiles and fits using the fuzzy sphere form factor.}
	\label{fig:fuzzy}
\end{figure}

Figure \ref{fig:fuzzy} and table \ref{tab:FuzzyParameters} show the results of fitting with the fuzzy sphere model\cite{stieger2004small} the SANS intensity profiles. The fuzzy sphere model is given by
\begin{equation}\label{eq:fuzzy}
   P(q)= \frac{3\left[ \sin{(qR)} - qR\cos{(qR)} \right]}{(qR)^{3}}\exp{\left( \frac{-(\sigma q)^{2}}{2} \right)}
\end{equation}
The fuzzy sphere model is unable to fit the intensity profiles, specially at the medium to large $q$ values.
The unsuccessful results of this model led us to choose the star polymer model as described in the manuscript.

\begin{table}[h]
	\centering
	\begin{tabular}{|c|c|c|c|} 
				\hline
				 & 20°C &30°C & 40°C\\ 
				\hline \hline
				$R$ (\AA) & 700  & 650 & 650 \\  
				\hline
				$\sigma$ (\AA) &  100  & 80  &  25 \\  
				\hline 
	\end{tabular}
	\caption{Parameters of the fuzzy model of Eq.~\ref{eq:fuzzy} obtained by fitting the experimental intensity profiles.}
	\label{tab:FuzzyParameters}
\end{table}

\FloatBarrier
\section{Simulation Hertz Modulus}

\begin{figure}[t]
	\centering
	\includegraphics[width=1.0\textwidth]{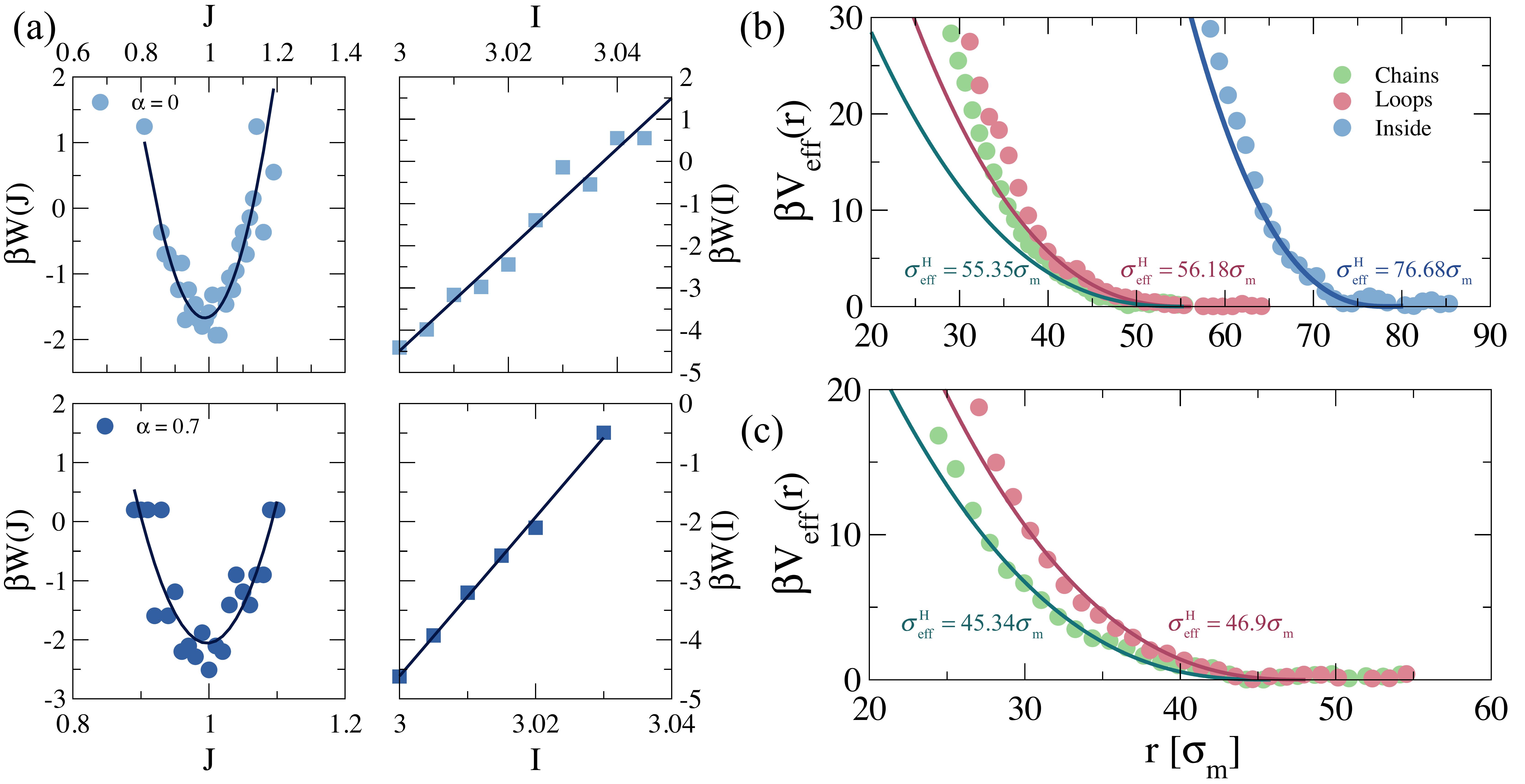}
	\caption{(a) Potential of mean force $W\left(J\right)$ and $W\left(I\right)$ for a composite microgel with polymer PEG chains inside at $\alpha=0$ and $0.7$. Effective potential interaction $\beta V_{eff}\left(r\right)$, where symbols are simulation results and the complete lines are fits with the Hertzian model considering elastic moduli at (b) $\alpha=0$ and (c) $\alpha=0.65$}
	\label{fig:Mech}
\end{figure}

Fig.~\ref{fig:Mech}(a) shows the potential mean force $W\left(J\right)$ and $W\left(I\right)$ for a composite microgel with PEG chains distributed inside of the PNIPAM microgel at $\alpha=0$ and $\alpha=0.7$, with the respective fitting. In Fig.~\ref{fig:Mech}(b) and (c), we show the fit of the effective potential interaction computed by umbrella sampling $\beta V_{eff}$  with the Hertzian model defined as

\begin{equation}
    \beta V_{H}\left(r\right)=U=\frac{2Y{\sigma_{eff}^{H}}^{3}}{15\left(1-\nu^{2}\right)}\left(1-\frac{r}{\sigma_{eff}^{H}}\right)^{5/2}\theta\left({\sigma_{eff}^{H}-r}\right),
\label{eq:Hertfit}
\end{equation}

\noindent Here, the effective Hertzian particle diameter $\sigma_{H}$ is a fitting parameter.

\clearpage
\bibliography{report} 
\bibliographystyle{spiebib} 